# The Wide Field Monitor (WFM) of the China-Europe eXTP (enhanced X-ray Timing and Polarimetry) mission

Margarita Hernanz*[a,b], Marco Feroci[c], Yuri Evangelista[c], Aline Meuris[d], Stéphane Schanne[d], Gianluigi Zampa[e], Chris Tenzer[f], Jörg Bayer[f], Witold Nowosielski[g], Malgorzata Michalska[g], Emrah Kalemci[h], Müberra Sungur[i], Søren Brandt[j], Irfan Kuvvetli[j], Daniel Alvarez Franco[a,b], Alex Carmona[a,b], José-Luis Gálvez[a,b], Alessandro Patruno[a], Jean in' t Zand[k], Frans Zwart[k], Andrea Santangelo[f], Enrico Bozzo[l], Shuang-Nan Zhang[m], Fangjun Lu[m], Yupeng Xu[m], Riccardo Campana[n], Ettore Del Monte[c], Francesco Ceraudo[c], Alessio Nuti[c], Giovanni Della Casa[c], Andrea Argan[o], Gabriele Minervini[o], Matias Antonelli[p], Valter Bonvicini[p], Mirko Boezio[p], Daniela Cirrincione[p], Riccardo Munini[p], Alexandre Rachevski[p], Andrea Vacchi[p], Nicola Zampa[p], Irina Rashevskaya[q], Francesco Ficorella[r], Antonino Picciotto[r], Nicola Zorzi[r], David Baudin[d], Florent Bouyjou[d], Olivier Gevin[d], Olivier Limousin[d], Paul Hedderman[f], Samuel Pliego[f], Hao Xiong[f], Rob de la Rie[k], Phillip Laubert[k], Gabby Aitink-Kroes[k], Lucien Kuiper[k], Piotr Orleanski[g], Konrad Skup[g], Denis Tcherniak[j], Onur Turhan[i], Ayhan Bozkurt[h], Ahmet Onat[s]

[a]Institute of Space Sciences (ICE-CSIC), Campus UAB, 08193 Cerdanyola del Vallès (Barcelona), Spain; [b]Institut d'Estudis Espacials de Catalunya (IEEC), Barcelona, Spain; [c]INAF - Istituto di Astrofisica e Planetologia Spaziali, Roma, Italy; [d]IRFU, CEA, Université Paris-Saclay, 91191 Gif-sur-Yvette, France; [e]INFN-Trieste, 34127 Trieste, Italy; [f]Eberhard Karls Univ. Tübingen (EKUT), Germany; [g]Space Research Ctr. of the Polish Academy of Sciences, Poland; [h]Sabanci Univ., Turkey; [i]TÜBITAK Space Technologies Research Institute, Ankara, Turkey; [j]Technical Univ. of Denmark, Denmark; [k]SRON Netherlands Institute for Space Research, Netherlands; [l]Univ. de Genève, Switzerland; [m]IHEP -Institute of High Energy Physics, Beijing 100049, China; [n]INAF/OAS, Bologna, Italy; [o]INAF, Viale del Parco Mellini 84, I-00136 Rome, Italy; [p]INFN Trieste, 34149 Trieste, Italy; [q]TIFPA, Trento, 38123 Povo, Italy; [r]FBK – Fondazione Bruno Kessler, Povo (Trento), Italy; [s]Istanbul Technical University, Ayazağa Campus, Turkey

## ABSTRACT

The eXTP (enhanced X-ray Timing and Polarimetry) mission is a major project of the Chinese Academy of Sciences (CAS), with a large involvement of Europe. Its scientific payload includes four instruments: SFA (Spectroscopy Focusing Array), PFA (Polarimetry Focusing Array), LAD (Large Area Detector) and WFM (Wide Field Monitor). They offer an unprecedented simultaneous wide-band Xray timing and polarimetry sensitivity. A large European consortium is contributing to the eXTP study, both for the science and the instrumentation. Europe is expected to provide two of the four instruments: LAD and WFM; the LAD is led by Italy and the WFM by Spain. The WFM for eXTP is based on the design originally proposed for the LOFT ESA M3 mission, that underwent a Phase A feasibility study. It will be a wide field of view X-ray monitor instrument working in the 2-50 keV energy range, achieved with large-area Silicon Drift Detectors (SDDs), similar to the ones used for the LAD but with better spatial resolution. The WFM will consist of 3 pairs of coded mask cameras with a total combined field of view (FoV) of 90x180 degrees at zero response and a source localisation accuracy of ~1 arc min. The main goal of the WFM onboard eXTP is to provide triggers for the target of opportunity observations of the narrow field of view instruments (SFA, PFA and LAD), in order to perform the core science observation programme, dedicated to the study of matter under extreme conditions of density, gravity and magnetism. In addition, the unprecedented combination of large field of view and imaging capability, down to 2 keV, of the WFM will allow eXTP to make important discoveries of the variable and transient X-ray sky, and provide X-ray coverage of a broad range of astrophysical objects covered under 'observatory science', such as gamma-ray bursts, fast radio bursts, gravitational wave electromagnetic counterparts. In this paper we provide an overview of the WFM instrument, explaining its design, configuration, and anticipated performance. Right now, eXTP is in phase B2, after a successful I-SRR (Instrument System Requirements Review). It is waiting for the adoption of the whole eXTP mission in China. Details about the current work in Phase B2, including the manufacturing and testing of the demonstration models of the WFM subsystems, will be presented, paying also a special emphasis on the collaboration with space dedicated industrial partners.

**Keywords:** X-ray timing, X-ray spectroscopy, Coded mask imaging, Silicon Drift Detectors, Compact stars: black holes, neutron stars, white dwarfs. Gamma-ray bursts, X-ray bursts, Gravitational wave events.

*hernanz@ice.csic.es; phone +34 93 737 97 88, www.ice.csic.es

## 1. INTRODUCTION

The enhanced X-ray Timing and Polarimetry (eXTP) mission is a planned flagship X-ray observatory, led by China, for X-ray timing, spectroscopy and polarimetry. Europe has been expected to provide two of its four instruments: the LAD (Large Area Detector), led by Italy, and the WFM (Wide Field Monitor), led by Spain. The contributions to the WFM of eXTP from Europe, in addition to Spain, are from Italy, France, The Netherlands, Germany, Poland, Denmark and Turkey. Regarding the LAD, in addition to Italy there are contributions from France, Germany, Poland, Czech Republic and Austria. This set of instruments offers a unique simultaneous wide-band (0.5 - 50 keV) with X-ray spectral, timing and polarimetric capability. A full description of eXTP can be found in the review papers [1, 2, 3].

The two Chinese instruments from the original eXTP include 13 focusing mirrors, 9 in the SFA and 4 in the PFA. The LAD instrument, led by Italy, includes 40 modules, whereas the WFM is composed of 6 cameras. See Figure 1 (left) for an artist impression of the original eXTP satellite with all its four instruments. (See the paper about the LAD - Feroci et al.- in these same proceedings for more details about that instrument).

The main scientific goals of eXTP are related to very relevant topics of Fundamental Physics, with still open questions to answer, like understanding the nature of matter under extreme conditions of density (Equation of State - EOS - of Neutron Stars) and gravity (Strong Field Gravity, in the vicinity of Black Holes) as well as the propagation of light in extremely strong magnetic fields (Strong Magnetism). A full description of the three big topics of the eXTP "Core Science" mentioned above can be found in the review papers [4, 5, 6].

In the course of the process for the final mission adoption of the eXTP mission, the Chinese Academy of Sciences (CAS) required the removal of all programmatic uncertainties for the mission implementation, the largest residing in securing the European provision of the LAD and WFM payloads. As a result of this process, the LAD and WFM instruments have been shifted to the optional payloads for eXTP, which at the time of writing only includes 2 scientific payloads, the SFA and the PFA. In the same process, the rescoped mission was reduced in size and cost, leading to a design with 6 telescopes for the SFA (in place of the original 9) and 3 telescopes for the PFA (in place of the original 4). The design of the individual SFA and PFA telescopes remains the same. The new configuration leads to a reduction of mass, from 5.3 tons to 4.2 tons, thus fitting the smaller launcher CZ-3BE in place of the CZ-5 requested for the full-scale mission. Also, the target orbit was significantly changed, passing from the original low-Earth equatorial orbit to a baseline high-Earth orbit with 5.000 km perigee and 165.500 km apogee. The target launch date remains no later than 2030. The current design of eXTP is shown in Figure 1 (right), compared to the original design, shown also in Figure 1 (left), that was the context in which the study of the WFM instrument presented in this paper was carried out.

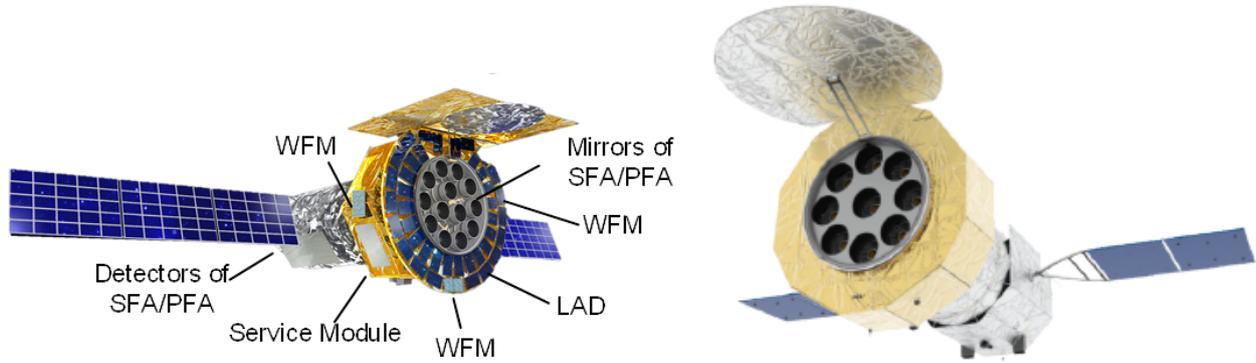

Figure 1. (Left): Artist impression of the original configuration of the eXTP mission with all its four instruments: SFA and PFA (led by China), LAD and WFM (led by Europe). (Right): Today's baseline design of eXTP, including only the SFA and PFA instruments.

## 2. THE SCIENTIFIC REQUIREMENTS OF THE WFM

The main scientific goals of the WFM are:

- Alert about the outbursts of the relevant sources - new or recurrent already known.
- Provide triggers for the "narrow field of view" instruments of eXTP: SFA, LAD, PFA: accreting black holes transients, accreting neutron stars transients.

Therefore, there's the need of an instrument with a Field of View (FoV) as wide as possible, to monitor the sky for the X-ray transients that the "Narrow Field of View" (NFI) instruments, LAD, SFA and PFA, should observe in depth.

In addition to providing the alerts for the LAD, SFA and PFA instruments, the WFM is able to do its own science, the so-called "Observatory Science". Some of its additional goals are:

- Monitor the long-term behavior of interesting X-ray sources.
- Detect short (0.1-100 s) bursts and record data with full resolution.
- Provide GRB (gamma-ray burst) alerts: quick dissemination of GRB sky positions via Beidou (Chinese GPS) .
- Provide GW (gravitational wave events) EM (electromagnetic) alerts: quick dissemination of GW EM searches after uplink of GW location.
- Provide alerts for FRB (Fast Radio Bursts): same as for GRBs and GWs (TBC).

A complete description of the Observatory Science that would be possible with the WFM of eXTP can be found in the review paper [7].

The scientific objectives of the WFM – listed as requirements and goals for each item - are listed in Table 1 (see also the eXTP/WFM papers [8, 9]).

Table 1: Scientific requirements of the WFM instrument

| Item | Requirement | Goal |
|---|---|---|
| Point source localization (confidence level 90%) | ≤ 1 arcmin | ≤ 0.5 arcmin |
| Angular resolution | ≤ 5 arcmin | |

| Item | Requirement | Goal |
|---|---|---|
| Peak sensitivity (5 σ source detection) | 1 Crab (1 s) <br> 5 mCrab (50 ks) | 0.8 Crab (1s) <br> 4 mCrab (50 ks) |
| Absolute flux calibration accuracy | 20 % | 15 % |
| Relative flux calibration precision | 5 % | 2.5 % |
| Sensitivity variation knowledge | 10 % | 5 % |
| Duration for rate triggers | 0.01 - 100 s | 1ms - 100 s |
| Field of view | ≥ 3.1 sr (around all the pointing instruments) | 1.75 π = 5.5 sr at zero response / 1.33 π = 4.2 sr at 20% of peak camera response |
| Energy range | 2 – 50 keV primary | 1.5 – 50 keV primary |
| Energy resolution | ≤500 eV | ≤300 eV |
| Energy scale calibration accuracy | ≤4% | ≤2% |
| Number of energy bands for compressed images | ≥8 | ≥16 |
| Time resolution | 300 sec for images <br> 10 μsec for event data | 150 sec for images <br> 5 μsec for event data |
| Absolute time calibration accuracy | 2 μsec | 1 μsec |
| Rate meter data | 16 ms | 8 ms |
| Event/image data downlink maximum delay | 3 hours | 1.5 hours |
| Onboard storage of triggered date | 3 hours | |
| Broadcast of trigger time and position to end user | ≤30 sec for 65% of the events after on-board detection of the event | ≤ 20 sec for 75% of the events |
| Number of triggers for WFM | ≥5 alerts per day | ≥ 1 per orbit |
| Modularity | No full loss of FoV due to single point failure | |
| On-board memory | 5 min @ 100 Crab | 10 min @ 100 Crab |

### 3. FUNCTIONAL DESCRIPTION AND OPTICAL DESIGN

In order to fulfill the scientific goals described in section 2, the WFM has been designed to have the capability to simultaneously observe about 1/3 of the sky (1.33*pi sr =33% at 20% of peak response, and 1.75*pi=44% at zero response) in the same energy band as the LAD instrument. Since the main purpose of the WFM is to detect sources for follow-up observations with the LAD, SFA and PFA, its field of view is designed to have a maximum overlap with the sky accessible to LAD, SFA and PFA pointing. These sources are new transients, as well as known sources undergoing spectral state changes.

The WFM should be an imaging instrument with a point source localization accuracy of 1 arc minute, to match the required on-target pointing accuracy for the LAD and the other NFI instruments observations. Its conceptual design is based on the classical coded mask design, successfully employed by several instruments on past and on-going X-ray missions, like SIGMA on GRANAT. WFC on Beppo-SAX, SPI, IBIS and JEM-X on INTEGRAL, SuperAGILE on AGILE, BAT on SWIFT. Also, the WFM should have the capability to simultaneously observe about 1/3 of the sky in the same energy band as the LAD, in order to fulfill its scientific objectives.

The development of the WFM for eXTP is based on the design originally proposed for the LOFT ESA M3 mission (see [10]) that underwent a Phase A feasibility study. A LAD instrument similar to the one of eXTP was also part of the LOFT proposal payload.

The WFM instrument has a modular design with six identical cameras, as will be explained below. Each one of the cameras is a coded aperture imaging device working in the X-ray energy range. Its working principle is the classical sky encoding by coded masks, widely used in X- and gamma-ray spatial instruments. The mask shadow recorded by the position-sensitive detector can be deconvolved to recover the image of the sky, with an angular resolution given by the ratio between the mask element and the mask-detector distance.

The design of each WFM camera is explained schematically in Figure 2. Photon detection is based on large-area Silicon Drift Detectors (SDDs), the same used for the LAD but with a smaller anode pitch (169 μm in the WFM versus 970 μm in the LAD). In this way, imaging properties are optimized, when SDDs are combined with the coded mask. When a photon is absorbed in the SDD, a charge - electron cloud - is generated, and it drifts towards the anodes. The size of the charge cloud increases during the drift and its size can be measured, thanks to the small anode pitch, that permits to record it on more than one anode (see Figure 2, right panel). This size gives information of the vertical location of the photon impact, whereas the anode position provides the horizontal location. Therefore, the SDD detectors provide an asymmetric imaging capability: the photon impinging on the detector will have fine position resolution in the anodes-direction (better than ~60 μm) and coarse resolution in the drift (perpendicular) direction (better than ~8 mm), as shown in the right panel of Figure 2. The coded mask pattern has been chosen to match with the asymmetric SDD (see middle panel of Figure 2). As also seen in Figure 2 (left panel), each camera includes four large area SDDs.

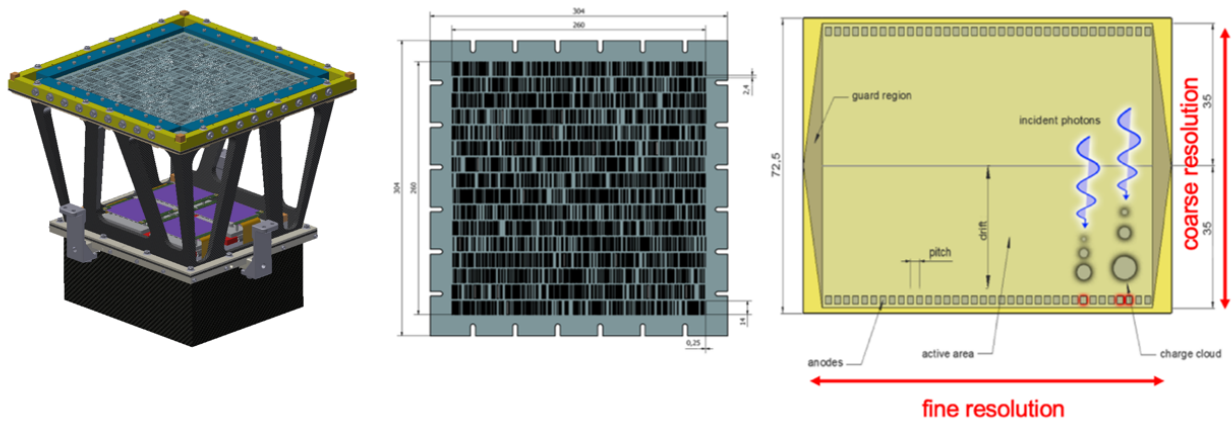

Figure 2. WFM camera concept. Right: scheme of the SDD photon working principle, with indication of the fine (anodes) and coarse (drift) resolution directions. The pitch of the SDD anodes is 169 μm. Middle: coded mask, with the long slit (coarse resolution) and short slit (fine resolution) directions oriented as in the detector. Left: WFM camera with the detector plane - including four detector assemblies (blue color) - and the coded mask mounted on top, taking into account the fine and coarse resolution directions. The collimator structure and the back-end electronics box below the detector plane are also shown.

By observing simultaneously the same sky region with two cameras oriented perpendicularly to each other (those two cameras are named a WFM camera pair), the precise 2D position of the sources can be derived, by intersecting the two orthogonal 1D projections. This WFM two-dimensional (2D) imaging principle is shown for a point source in Figure 3.

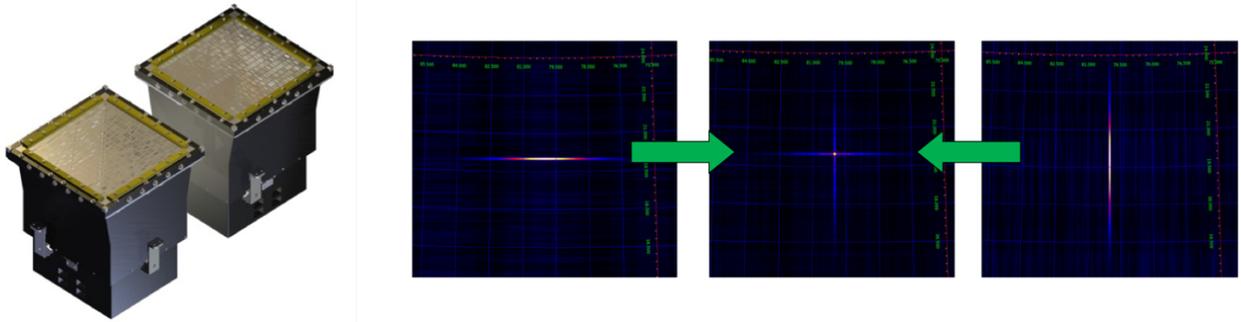

Figure 3. WFM 2D imaging principle. Left: WFM camera pair, where two identical cameras are arranged orthogonally (regarding their coded mask orientations). Right: the combination of the two images obtained with each one of the two cameras of a WFM camera pair (right and left figures) gives the 2D image (central panel). This simulated image corresponds to an isolated source.

The optical configuration of each WFM camera is shown in Figure 4 and detailed in Table 2. With such configuration, the detector's (SDDs) and coded mask properties, both the imaging (e.g., 1 arcmin source localization accuracy in two dimensions) and the energy (range and resolution) requirements are fulfilled. Then, each camera has a field of view 90ºx90º (full width at zero response) and 30ºx30º (fully coded, i.e., with fully illuminated detectors).

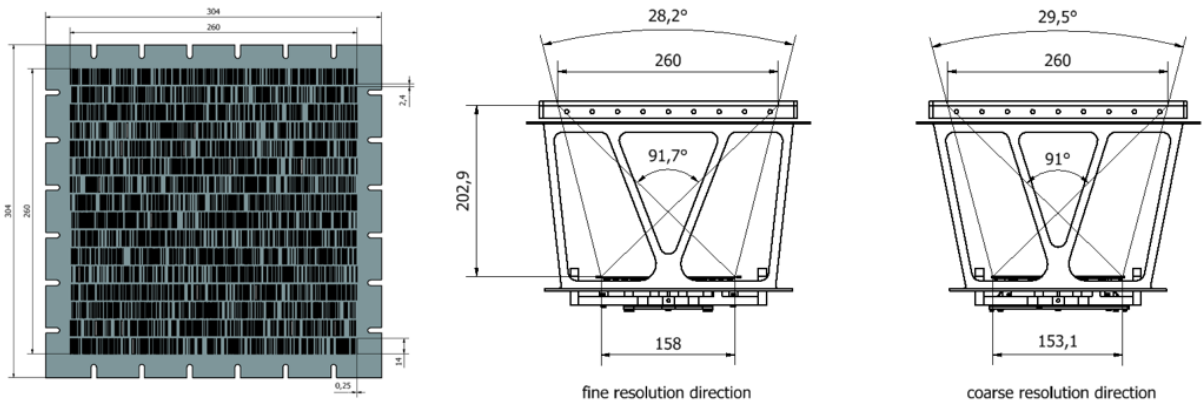

Figure 4. Optical configuration of the WFM cameras: coded mask pattern and scheme of each camera with dimensions.

Table 2: Optical design and performance of the WFM instrument

| Item | Value |
|---|---|
| WFM cameras | 6 (arranged in 3 camera pairs) |
| **Optical design - Coded mask properties** | |
| Mask pattern pitch (pixel size) | 250 $\mu$m x 16.4 mm |
| Size of mask open (slits) and closed elements | 250 $\mu$m x 14 mm |
| Mask size | 260 mm x 260 mm$^2$ x 150 $\mu$m (thickness) |
| Mask-detector distance | 202.9 mm |
| Open mask fraction (effective value) | 25 x 0.85 = 21% |

| Item | Value | | |
|---|---|---|---|
| **Optical design - Detector properties** | | | |
| Detector type - number of SDD tiles | Si Drift (SDD) - 24 (4 per camera) | | |
| Detector spatial resolution (FWHM) | < 60 $\mu$m (fine direction); < 8mm (coarse direction) | | |
| Detector size (single) | 77.4 mm x 72.5 mm x 450 $\mu$m | | |
| Detector active area (single) | 64.9 mm x 70.0 mm = 4543 mm$^2$ = 45.43 cm$^2$ | | |
| | **Camera** | **Camera pair** | |
| WFM detectors effective area | 182 cm$^2$ | 364 cm$^2$ | |
| WFM peak effective area (on axis, through mask) | > 38 cm$^2$ | > 76 cm$^2$ | |
| **Optical performance** | | | |
| | **Camera** | **Camera pair** | **WFM** |
| Angular resolution | < 5 arcmin x 5º | < 5 arcmin x < 5 arcmin | < 5 arcmin |
| Point Source location accuracy (PSLA; SNR> 10 $\sigma$, | 0.9 arcmin x 30 arcmin | 0.9 arcmin x 0.9 arcmin | 0.9 arcmin |
| Field of view: full width at zero response (FWZR) | 90º x 90º | 90º x 90º | 180º x 90º |
| Field of view: fully illuminated detectors | 30º x 30º | 30º x 30º | 90º x 30º |

## 4. ACCOMMODATION ON THE SPACECRAFT

The six cameras of the WFM arranged in three camera pairs and accommodated on the spacecraft are shown in Figure 5. The three camera pairs are oriented to cover the required large fraction of the sky: the camera pair with cameras 0A and 0B points to the viewing direction of the "Narrow Field of View" (NFI) instruments LAD, SFA and PFA. The two other camera pairs (cameras 1A-1B and 2A-2B) are tilted +/- 60° relative to that direction. With such configuration, the wide field of view required by the scientific requirements (see Table 1) is fulfilled. And it is important to point out that the 6 cameras are identical and can be placed anywhere in the satellite platform, provided that they keep their orientation, However, the field-of-view (FoV) of the WFM should not be affected by the other components placed on the optical bench (antennae, star-trackers, Sun shade), and the WFM has to be placed in the shadow of a Sun shade. The WFM thus succeeds to have an exceptional and unprecedented large FoV (Figure 6), together with an excellent imaging capability in the energy range (2-50) keV, thanks to the coded mask design, the excellent spatial resolution of the SDD detectors and the configuration and orientation of the six identical cameras.

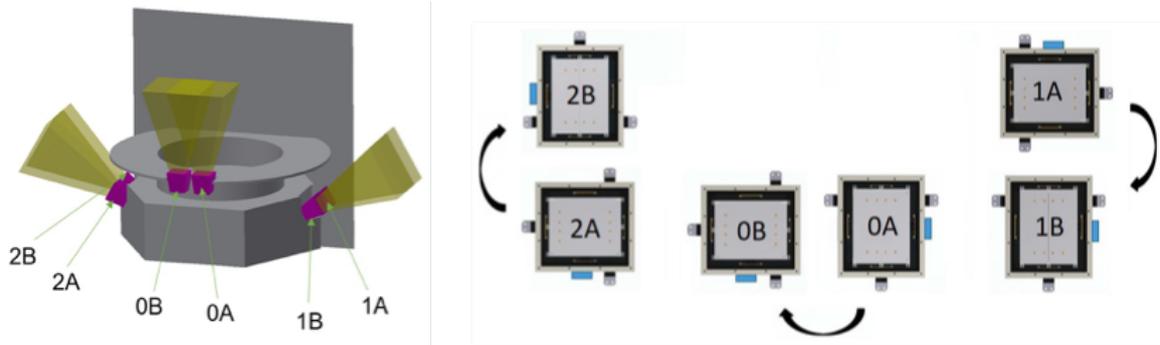

Figure 5. Accommodation of the six WFM cameras, organized in three camera pairs, in the eXTP satellite platform.

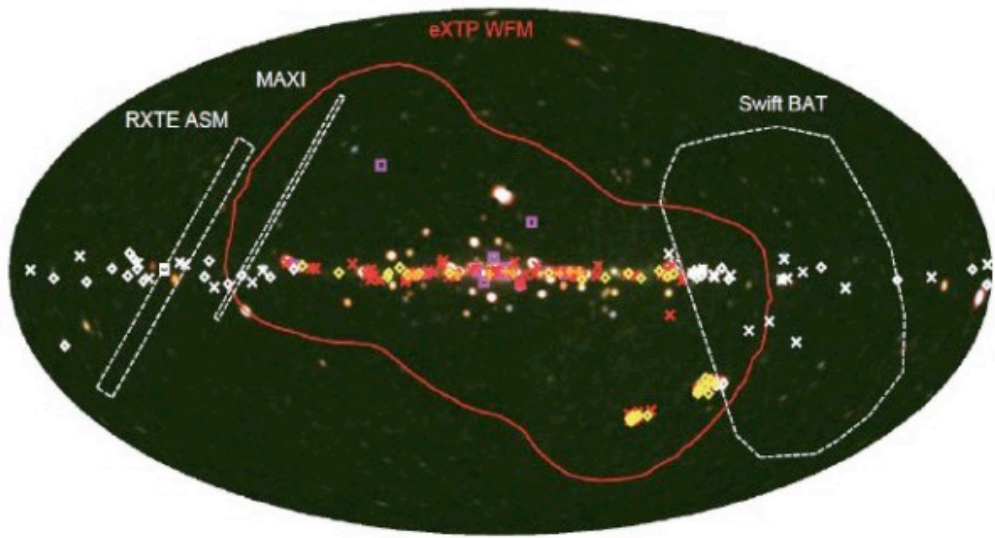

Figure 6. WFM instantaneous field of view, compared with the most relevant existing facilities (background courtesy of T. Mihara, RIKEN, JAXA and the MAXI team).

## 5. WFM FUNCTIONAL DESIGN

The whole WFM is composed of three camera pairs (six cameras) and two Instrument Control Units (cold redundancy). Each camera is composed of a detector tray with four Silicon Drift Detectors, four Front-End Electronics, four Be (or equivalent material) protecting windows (of the detectors), a Back-End Electronics assembly, a Collimator assembly, and a Coded Mask assembly with a thermal foil. The camera pairs and the cameras in the WFM are organized to achieve a high level of redundancy. The functional block diagram for the WFM is shown in Figure 7.

The main functions of the WFM FEE are:
- Forward filtered bias voltages to the SDD
- Forward power and configuration data to the ASICs (but in fact the ASICs are an integral part of the FEE and thus don't need to be specified here)
- Read-out the SDD signals
- Interface the Back-End Electronics
- Mechanical support for the SDD

The main functions of the BEE are:
- A/D convert the SDD signals

- Time tagging of the X-ray events
- Trigger selection
- Pedestal subtraction
- Common mode noise subtraction for the fit of the X-ray signal
- Determination of charge cloud center and width (yielding position in the fine and coarse direction)
- Reconstruction of the total charge collected, correcting for channel gains

The main functions of the WFM ICU are:
- Interfacing the BEEs
- TC and configuration handling
- On board time management
- Image integration
- Burst triggering

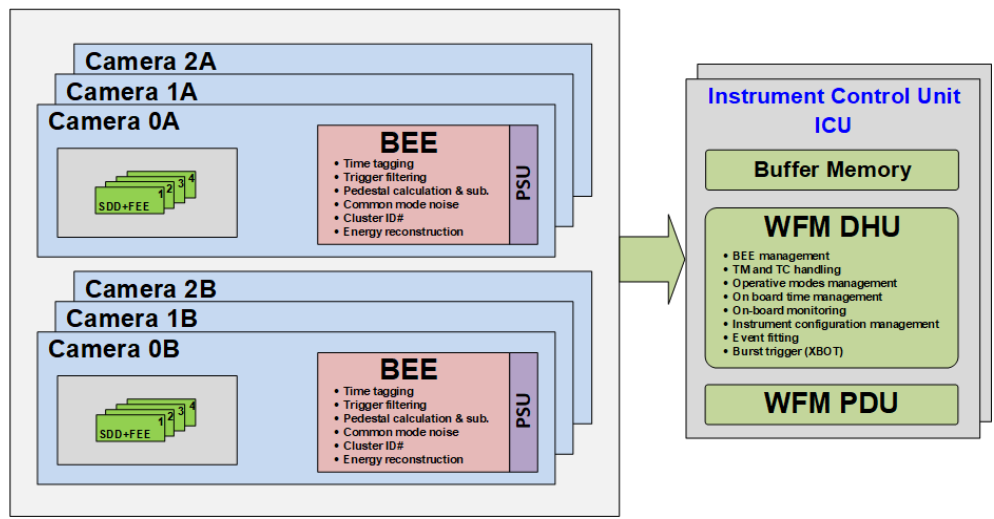

Figure 7. Functional block diagram of the WFM instrument, including the 3 camera pairs, and the main and redundant Instrument Control Units (ICUs).

# 6. WFM PRODUCT TREE

The different subsystems of the WFM, together with their European responsible teams, are shown in the Product Tree (Figure 8).

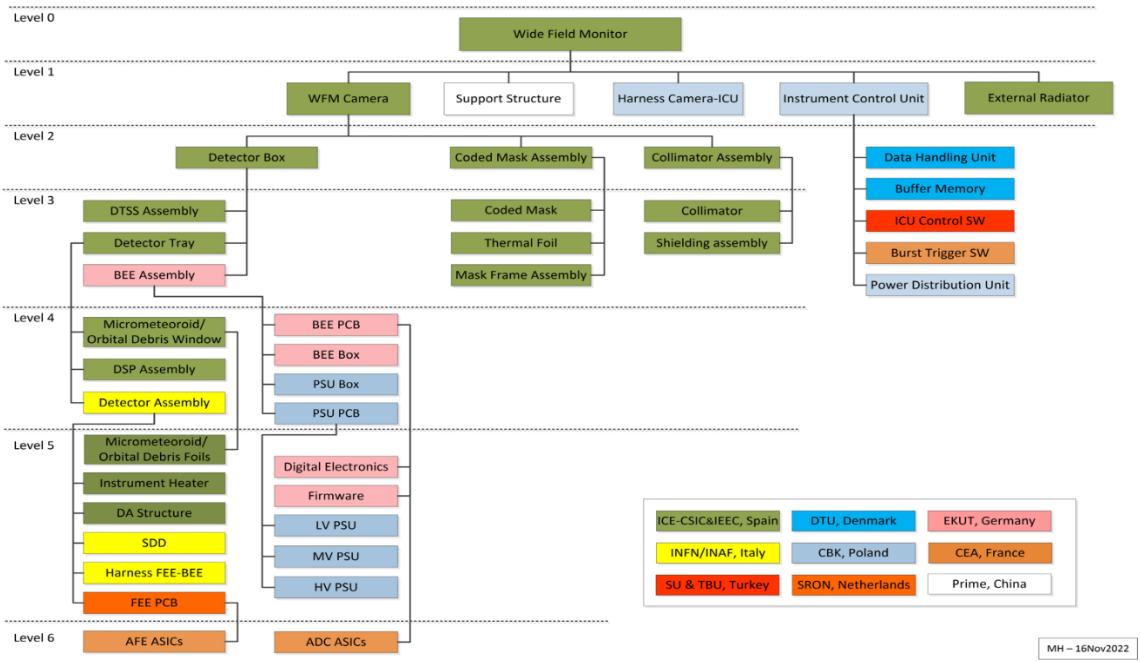

Figure 8. Product Tree of the WFM for eXTP.

## 7. WFM CAMERA DESIGN

The layout of a WFM camera and its exploded view are shown in Figure 9.

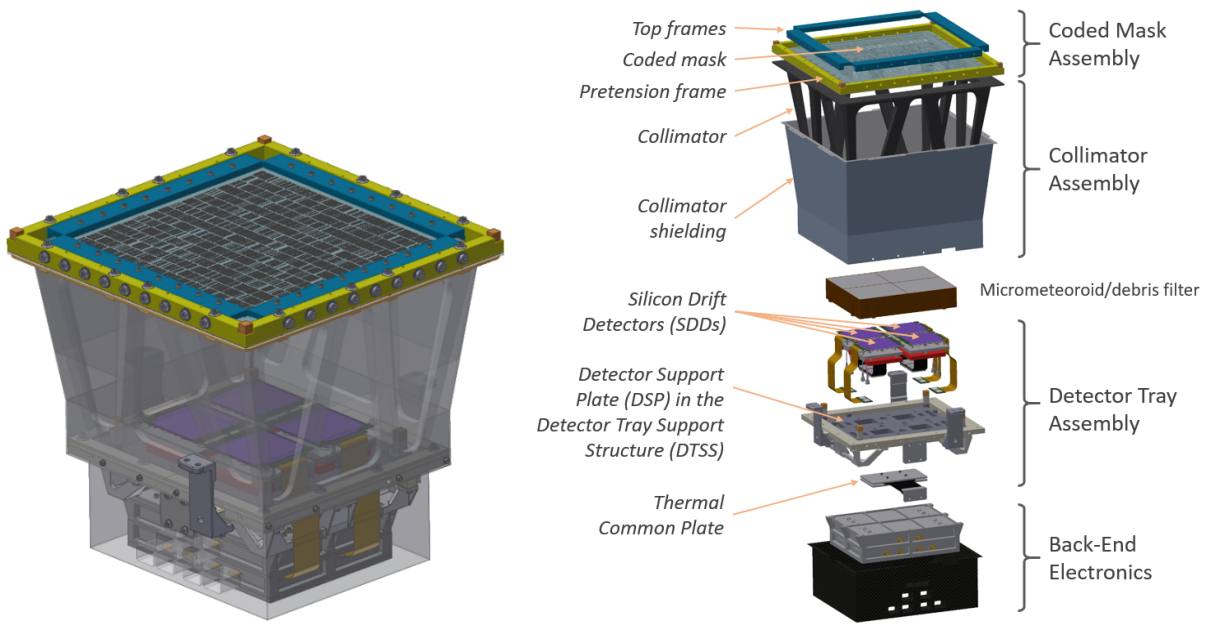

Figure 9. Layout of the WFM camera (left). Exploded view of the WFM camera, with its main components (right).

The coded mask assembly of each one of the six cameras includes the coded mask itself and the coded mask frames (see WFM Product Tree in Figure 8). Due to the small thickness of the mask (0.15 mm), it is necessary to keep it under tensile stress to avoid buckling. For this reason, a frame set acts as a pretension mechanism in order to minimize the vertical displacement of the mask in the operational mode and to provide enough stiffness for surviving the launch. Details of the optical properties of the mask are included in Table 2. On top of the coded mask there is a thermal foil, aimed to reduce as much as possible the mask temperature variations along the orbit, in order to keep the mask flatness as constant as possible.

The collimator structure supports the coded mask frame assembly and is made of CFRP structure, that will permit to have access to the detector tray after the assembly of the camera – for alignment purposes - and will have enough stiffness to avoid any deformations which can appear during launch (accelerations) and operation of the WFM (thermal stresses).

A detailed description of the mechanical structure of the WFM can be found in the SPIE2022 paper "The mechanical design and implementation of the WFM cameras for eXTP" [14]

As shown in the WFM Product Tree (Figure 8), the Detector Box consists of the following subsystems:

- Detector Tray Support Structure (DTSS)
- Detector tray
- Back-End Electronics (BEE) assembly

The Detector Tray includes:
- 4 Detector Assemblies (DAs)
- A Micrometeoroid / Orbital debris Window (also called Beryllium - Be- Window, although it can be made of an alternative material, like Polypropylene)
- A Detector Support Plate (DSP) accommodating the 4 DAs

The DA itself includes:
- SDD detector
- Readout Front-End Electronics (FEE) PCB - including the analog ASICs. (The ADC ASICs are physically located inside the BEE box)
- Harness FEE-BEE
- Instrument heater
- Mechanical structure: Cooling Plate (CP), Invar Bracket (IB)

The BEE Assembly includes:
- BEE box (mechanical structure)
- BEE data processing PCB, with its electronic components and Firmware
- Power Supply Unit (PSU): Low Voltage (LV), Medium Voltage (MV) and High Voltage (HV) PSUs

Detailed descriptions of the design and development of the detector assembly DA (including the detector/readout-electronics), the digital data processing concept (BEE) and the Instrument Control Unit processing hardware and software (ICU) of the WFM can be found in the Proc. SPIE 12181 (2022) [11], [12] and [13] papers.

## CURRENT STATUS

In 2023, both the LAD and the WFM (European instruments of eXTP) successfully passed the I-SRR (Instrument Systems Requirements Review), thus ending the phase B1.

Below we show pictures of some of the demonstration models of the WFM subsystems that have been developed during phase B.

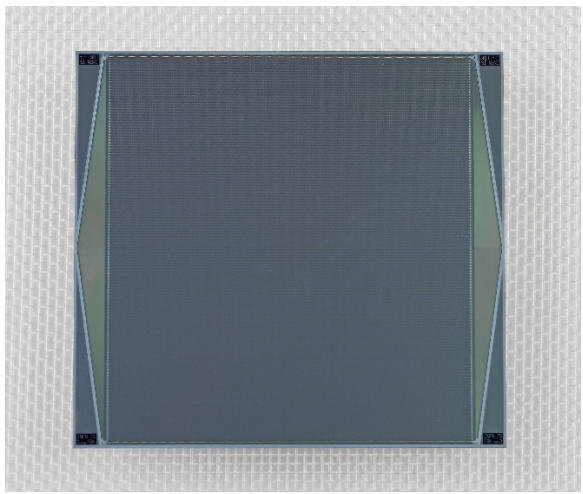 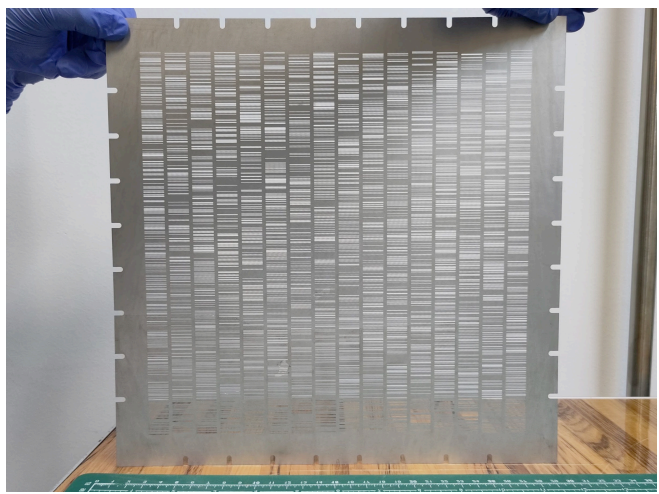

Figure 10. Left: Full size SDD detector for the WFM, designed and manufactured by INFN, INAF and FBK (from INFN & FBK). Right: Full size coded mask with final pattern (CSIC)

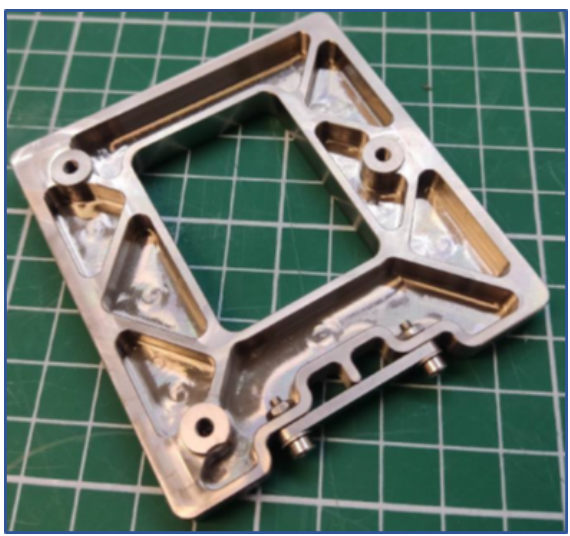 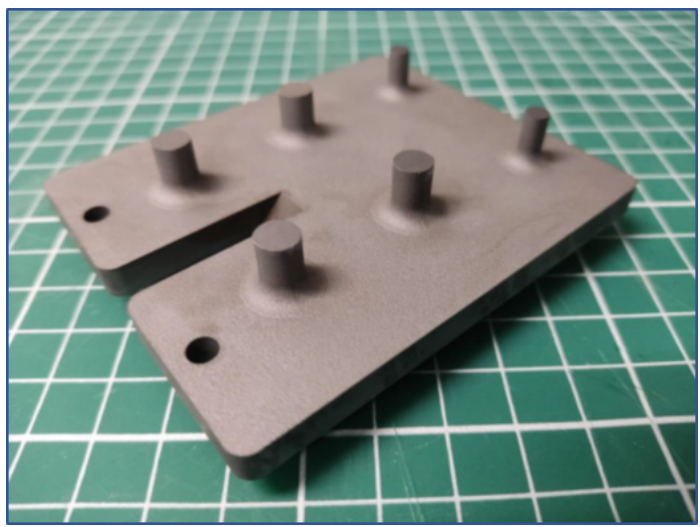

Figure 11. Invar Bracket (left) and Cooling Plate (right) from CSIC provided to SRON for the whole Detector Assembly manufacturing. These are the mechanical and thermal interfaces of the WFM-detector sandwich (SDD plus FEE) with the DA and the detector tray.

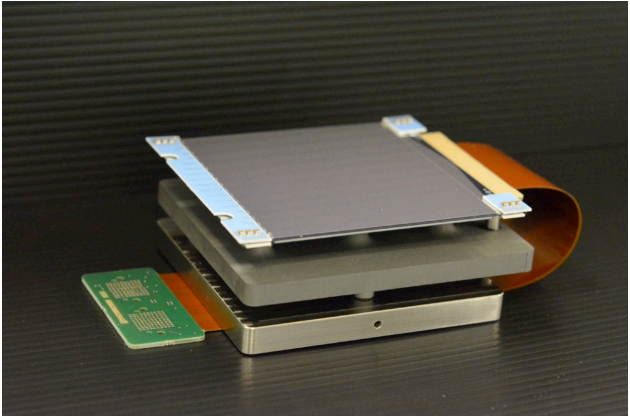
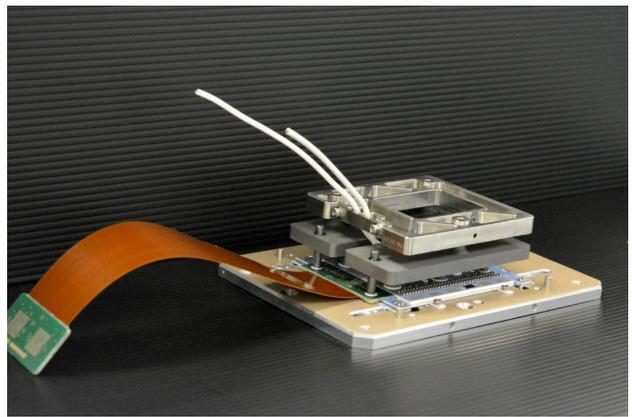
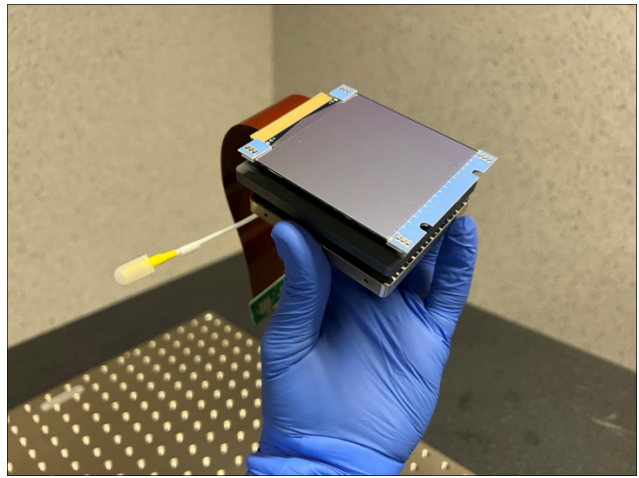

Figure12. DA dummy to be used for the alignment tests of the DAs in the detector tray: top view (top panel, left), bottom view (top panel right), general view (lower panel).

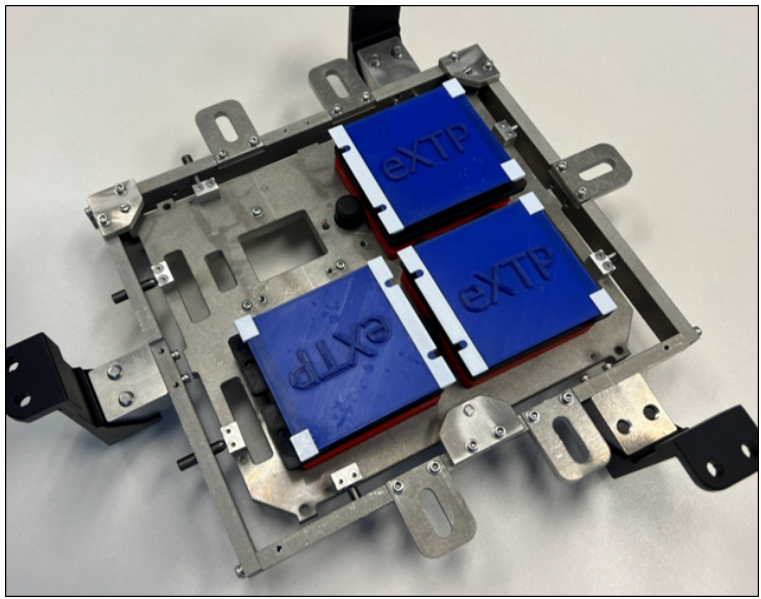

Figure 13. Mock-up of the DTSS (Detector Tray Support Structure), with the DSP (Detector Support Plate) and 3 DAs.

## CONCLUSIONS

The WFM design for the eXTP mission is very mature. A Demonstration Model program is being developed, including end of summer of 2025. Full-scale prototypes of the critical elements, like the detector assembly and the coded mask.

The WFM project is still funded in Spain until end 2025, thus allowing to develop a full-scale mechanical model of the whole WFM camera, as well as to test the alignment of the detectors in the detector tray and of the detector tray with the coded mask.

The WFM experiment is no longer in the eXTP baseline (as is also the case for the LAD instrument), but discussions are still ongoing at agency level to evaluate the possibility of an European participation to the Chinese eXTP mission approved in China.

## ACKNOWLEDGEMENTS

The CSIC team acknowledges funding support from MICIN/AEI PID2019-108709GB-I00 and EU Next Generation funds (Recovery, Transformation and Resilience Plan) MRR_TECNOUNIVERSE23/008. The Italian authors acknowledge funding support by the Italian Space Agency. The Polish team acknowledges NCN grant 2019/35/B/ST9/03944.